\documentclass[prl,twocolumn]{revtex4}

\usepackage{graphicx}
\usepackage{dcolumn}
\usepackage{amsmath}
\usepackage {epsfig}

\def\tp{$\tau$-phase }
\def\bt{$B_{th}$}

\def\tb{$\tau$-AuBr$_{2}$}

\makeatother
\begin{document}
\title[Short Title]{Disorder-driven magnetic field-dependent
phases in an organic conductor}
\author{K. Storr$^1$, D. Graf$^2$, J. S. Brooks$^2$, L. Balicas$^2$,
C. Mielke$^3$, and G.C. Papavassiliou$^4$ }
\affiliation{$^1$Physics Department, Florida A\&M University,
Tallahassee, FL 32307 USA} \affiliation{$^2$National High Magnetic
Field Laboratory and Physics Department, Florida State University,
Tallahassee, FL 32310 USA} \affiliation{$^3$NHMFL/Los Alamos
National Laboratory, Los Alamos, NM 87545 USA}
\affiliation{$^4$Theoretical and Physical Chemistry Institute,
National Hellenic Research Foundation, Athens, 116-35, Greece}
\date{\today}

\begin{abstract}
We report inter-plane ($R_{zz}$) electrical transport measurements
in the \tp series of organic conductors at very high magnetic
fields. In the field range between 36 and 60 T $R_{zz}$ shows a
very hysteretic first order phase transition from metallic to an
insulating state. This transition does not affect the
Shubnikov-de-Haas oscillations associated with the two-dimensional
(2D) Fermi surface. We argue that this transition originates from
inter-plane disorder which gives rise to incoherent transport
along the least conducting axis. We conclude that this system
becomes a strictly 2D Fermi-liquid at high magnetic fields.
\end{abstract}

\maketitle

Disorder and complexity play important roles in many systems of
current interest in condensed matter physics. In colossal
magnetoresistance (CMR) systems\cite{cmr} such as the manganites,
magnetic disorder and inhomogeneity lead to dramatic changes in
resistivity with magnetic field and temperature\cite{tokura}. In
organic conductors with complex, polymeric anions and cation
end-groups, conformational disorder may lead to insulating
behavior, associated to the localization of the electron states,
and to weak ferromagnetism even in the absence of magnetic ions
\cite{welp,shus}.  As recent reviews have shown\cite{ross,elbio},
important mechanisms and physical properties are shared by organic
and inorganic materials, particularly in layered, anisotropic
systems. Recently ``bad metal" (i.e. $k_F\ell$$\approx 1$)
characteristics have been reported in a transition metal oxide
(TMO) that paradoxically, exhibits Shubnikov de Haas oscillations
($\omega_c\tau \gg 1$) at low temperatures\cite{cao}. This also
appears to be the case of the $\tau$-phase organic conductor
series presented in this Letter, where disorder, magnetism, and
high anisotropy play important roles in determining the ground
state properties of layered materials.
\begin{figure}[]
\epsfig{file=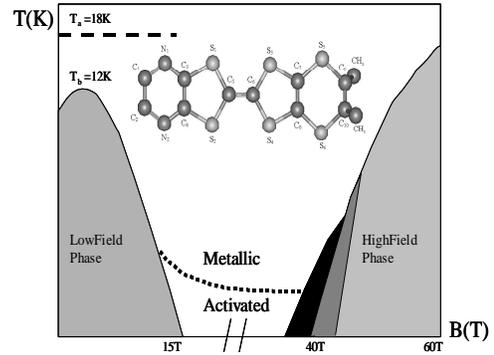,height=5cm,width=6.5cm} \caption{Schematic phase
diagram of the \tp organic conductors based on the DMEDT-TTF donor
molecule (inset). The low and intermediate field phase boundaries
are obtained from temperature dependent resistance measurements.
(See text for details.) Between 12 and 40 T quantum oscillations
characteristic of a 2D Fermi surface are observed. The threshold
fields B$_{th}$ for the high field phase boundaries are based on
pulsed field magnetoresistance data for (1) $\tau(r)$-AuBr$_{2}$,
(2) $\tau$-AuBr$_{2}$, (3) $\tau$-AuI$_{2}$, low to high,
respectively.}\label{fig1}
\end{figure}

Our main experimental result is the universal behavior of the
inter-plane resistance ($R_{zz}$) of three isostructural members
of the $\tau$-phase class of organic
conductors\cite{pap90,ter,zam,pap95,pap96} in very high ($B > 36
T$) magnetic fields. Here we find a first order metal-to-insulator
transition in $R_{zz}$ above a temperature-dependent threshold
field $B_{th}$. An overview of the field-temperature phase diagram
of these materials is shown in Fig. 1. What is remarkable about
this transition is that unlike the well-known field induced spin
density wave transitions in quasi-one dimensional organic
conductors\cite{ish}, where an orbital Fermi surface nesting
effect is involved\cite{osada}, $B_{th}$ appears to be only weakly
dependent on magnetic field direction. Hence the mechanism that
drives the transition at $B_{th}$ may be primarily isotropic, or
Zeeman-like in origin, and we must look for other physical
characteristics of these materials which cause this dramatic
transition to the insulating state.

The complexity and structure of the $\tau$-phase unit cell is
unique amongst the general class of charge transfer salts
(CTS)\cite{will}. Three forms of the $\tau$-phase materials have
been studied in the present work. The formula unit for the
material we have most extensively studied\cite{sto} (hereafter
$\tau$-AuBr$_2$) is
$\tau$-(P-(\emph{S,S})-DMEDT-TTF)$_2$(AuBr$_2$)(AuBr$_2$)$_y$
(where $y \approx 0.75$).  The iodine anion system
$\tau$-(P-(\emph{S,S})-DMEDT-TTF)$_2$(AuI$_2$)(AuI$_2$)$_y$
(hereafter $\tau$-(AuI$_2$) is also investigated in this work, as
is a disordered racemic system with AuBr$_2$(see below). The
DMEDT-TTF donor is shown in the inset of Fig. 1. Unlike
conventional CTS materials where the charge transfer is 2:1, here
it is 2:(1+y) and the anions occupy two different sites in the
unit cell: for instance the AuBr$_2$ anion sits at the center of a
square array of DMEDT-TTF donors in the conducting
layer\cite{zam}, and the (AuBr$_2$)$_y$ anions are arranged in the
inter-layer planes where their orientation alternates by 90$^o$
between successive layers. Since DMEDT-TTF is asymmetric, the
donor stacking involves alternating directions within the donor
layer. Hence, due to the very low symmetry of the donor and anion
arrangement, four donor layers are necessary to complete the unit
cell\cite{pap95}, and the result is an unusually large
inter-planar dimension: $(a,b,c \approx 7.4,7.4,68~
 \AA)$. \underbar{Further disorder}, beyond the non-stoichiometry of the anion
structure, can be introduced into the $\tau$-phase system with a
racemic mixture of two different isomers(\textit{R,R}
\underbar{and} \textit{S,S})~of the donor
molecules\cite{pap01a,pap01b}. In general the ratio of the two
isomers is not fifty-fifty, and this enhances the disorder over
crystals containing the pure (\textit{R,R} \underbar{or}
\textit{S,S}) arrangements. The racemic system studied in the
present work is
$\tau$-(P-(r)-DMEDT-TTF)$_2$(AuBr$_2$)(AuBr$_2$)$_y$ (hereafter
$\tau$(r)-AuBr$_2$).

    In this investigation, single crystals  of $\tau$-phase materials
(square plates of average size $1 \times 1 \times 0.2~mm^3$) were
grown by electrochemical methods. All measurements were
four-terminal, inter-plane resistance measurements with current
values between 50 nA and 300 $\mu A$. Electrical contact was made
to the samples with 25 micron gold wires connected with carbon or
silver paint. Experiments were carried out at the National High
Magnetic Field Laboratory (NHMFL) DC field facilities in
Tallahassee with standard low frequency AC techniques, and at
NHMFL-Los Alamos with pulsed magnets. For pulsed field
measurements, forward and reverse DC current traces were taken to
remove the induced \emph{emf}~ signals.

\begin{figure}[]
\epsfig{file=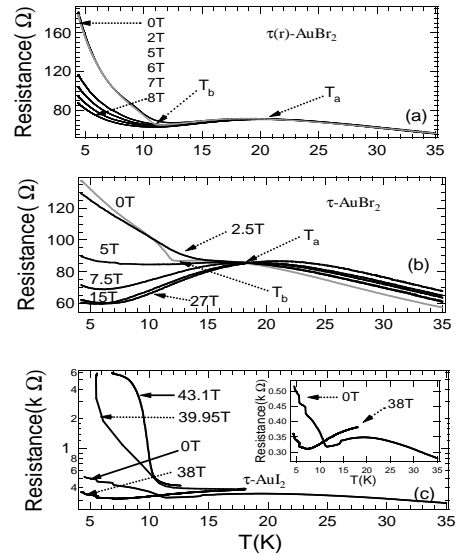,height=8cm,width=6.5cm} \caption{The temperature
dependence of the resistance of the three \tp materials for
different magnetic fields. ($T_a$ and $T_b$ are defined in the
text.)  a) $\tau$(r)-AuBr$_{2}$, b) $\tau$-AuBr$_{2}$, c)
$\tau$-AuI$_{2}$. Inset: expanded view of low resistance
traces.}\label{fig2}
\end{figure}

The temperature dependence of the inter-plane resistance for
different magnetic fields for the three \tp materials studied is
shown in Fig. 2. The following convention for presentation order
is used in the figures: $\tau-(r)-AuBr_2$, $\tau-AuBr_2$, and
$\tau-AuI_2$ to represent the level of disorder and/or anion radii
(where Br is smaller than I ) respectively. Although the data in
Fig. 2 differs in emphasis of field, temperature, and material
studied, there are universal features that arise in all cases, as
reflected in the phase diagram of Fig. 1. The inter-plane
resistance $R_{zz}$ increases monotonically with decreasing
temperature, but there is a characteristic temperature labelled
$T_a \approx 18 K$ where the resistance appears to be magnetic
field independent. (The magnetoresistance is positive above $T_a$,
and negative below $T_a$.) At lower temperatures an anomaly
(plateau or kink) appears in $R_{zz}$ for zero field at $T_b
\approx 12 K$. The detailed structure of the feature in $R_{zz}$
at $T_b$ quickly disappears with increasing magnetic field. If the
change of slope in the temperature dependence is monitored, $T_b$
first increases slightly, then decreases with increasing field. At
fields greater than 5 T, the point where dR/dT changes sign can be
easily determined. In all samples, at all fields, dR/dT eventually
becomes negative with decreasing temperature. Both $T_a$ and $T_b$
are shown in Fig. 1 as derived from these observations in the
field range 0 to 27 T.

\begin{figure}[]
\epsfig{file=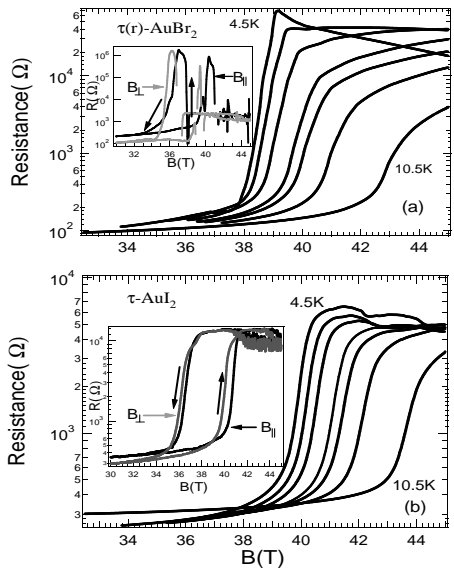,height=8cm,width=6.5cm}
\caption{Magnetoresistance studies of the high field transition in
the \tp materials in the hybrid magnet for different temperatures
(10.5 K, 9 K, 7.8 K, 7.1 K, 6.7 K, 5.6 K, 4.5 K). a)
$\tau(r)-AuBr_{2}$, b) $\tau-AuI_{2}$. Insets: Magnetoresistance
measurements for magnetic field parallel ( gray-scale:
$\theta$=90$^o$) and perpendicular (solid black: $\theta$ = 0$^o$)
to the conducting planes at T = 1.5 K. }\label{fig3}
\end{figure}

The high magnetic field dependence of $R_{zz}$ of the $\tau$-phase
materials has been investigated in a DC hybrid magnet up to 45 T (
Fig. 3), and in pulsed fields(6 ms rise-time) up to 60 T (Fig. 4).
In all cases a transition between a metallic state and a highly
insulating state is observed in $R_{zz}$ at a threshold field
$B_{th}$. In Fig. 5 details of the Shubnikov-de Haas (SdH) effect
are shown for representative low temperature hybrid data. (In the
present work SdH frequencies were observed over a range of 500 to
800 T, depending on field orientation. See also Ref.\cite{sto})
There are several key points that may be drawn from Figs. 2-5.
First is that between about 18 T and $B_{th}$ SdH oscillations are
observed at low temperatures even though $R_{zz}$ exhibits a
negative dR/dT in this activated region (Fig.1). Hence the sample
is a quasi-two dimensional metal in this range. Second, the
transition at $B_{th}$ is highly hysteretic, and therefore first
order. For fields above $B_{th}$, the inter-plane resistance
$R_{zz}$ increases by orders of magnitude, and in some cases
becomes un-measurable by conventional transport methods. $B_{th}$
is temperature dependent, and moves to higher fields with
increasing temperature. We note that the upper boundary of the
high field phase in Fig. 1 approaches a value near $T_a$. Third,
as Fig. 3 indicates, $B_{th}$ is only weakly dependent on field
orientation, and unlike the case for an orbital Fermi surface
nesting effect, $B_{th}$ actually reaches a \underbar{minimum} for
in-plane magnetic field.

Before discussing the mechanism for this new high field state, it
is instructive to consider some other details associated with the
$\tau$-phase materials. Even in the absence of magnetic field,
these materials have the appearance of ``bad metals". A simple
evaluation of the carrier scattering time, based on either
in-plane or inter-plane conductivity, over a broad temperature
range, leads to a mean free path comparable to the unit cell
dimensions (see also Ref. \cite{for}). In marked contrast, for
moderate magnetic fields where quantum oscillations (SdH) are
observed, we find a Dingle temperature\cite{sto} of only 1.3 K
which translates to a mean free path of about $60 \mu m$. This
difference is not easily reconciled based on any simple homogenous
model. That the disorder may arise from the inter-plane sites
comes from investigations of $R_{zz}$ in tilted magnetic
fields\cite{bro,sto} in \tb~ which show no evidence for the
``Yamaji effect"\cite{ish}. That is, no angular dependent
magnetoresistance oscillations(AMRO) in $R_{zz}$ associated with a
finite inter-plane bandwidth and mean free path are observed.
Hence the inter-planar coherence required for semi-classical
trajectories on a warped cylindrical Fermi surface is not
satisfied. Evidence for non-metallic behavior, coupled with the
appearance of quantum oscillations in quasi-two dimensional metals
are not without precedent, as recently reported in both
organic\cite{bal,wos} and TMO\cite{cao} systems.

\begin{figure}[]
\epsfig{file=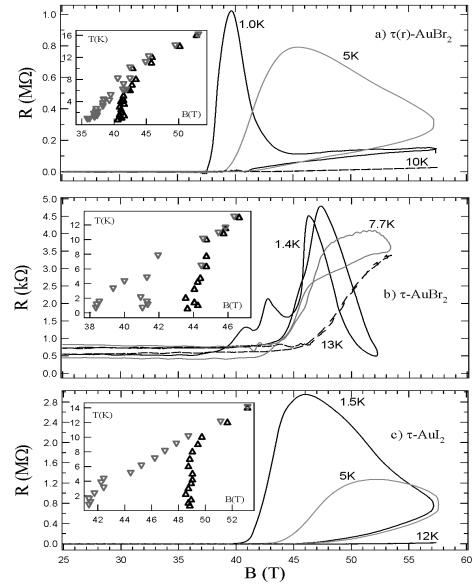,height=8cm,width=6.5cm}
 \caption{Magnetoresistance studies of the high field transition
in the \tp materials in pulsed magnetic fields. The loop-like
hysteresis in the signal in the high field phase is due in part to
the large RC time constants in the presence of a rapidly changing
magnetic field. a) $\tau(r)$-AuBr$_{2}$, b) $\tau$-AuBr$_{2}$,  c)
$\tau$-AuI$_{2}$. Insets: hysteretic high field phase boundaries
determined from linear extrapolation of the slope of $dR_{zz}/dB$
(at the onset of the insulating phase) to the field axis. In
$\tau$-AuBr$_{2}$  SdH oscillations are apparent, as is some
evidence for sub-phases, in the high field phase.}\label{fig4}
\end{figure}

\begin{figure}[]
\epsfig{file=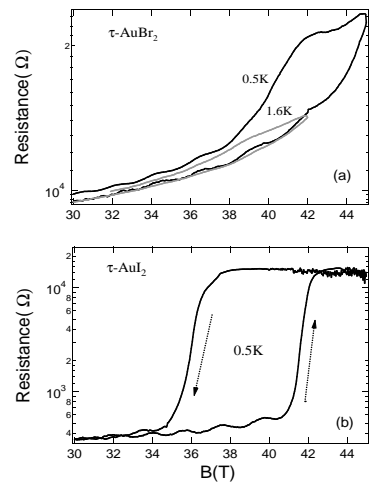,height=8cm,width=6.5cm} \caption{The SdH signal
in the vicinity of $B_{th}$ for a) $\tau$-AuBr$_2$ at T = 0.5 K up
to 45 T, and T = 1.6 K up to 42 T), b) $\tau$-AuI$_2$ at 0.5 K up
to 45 T . (The SdH signal for $\tau(r)$-AuBr$_{2}$ has not yet
been clearly observed.)}\label{fig5}
\end{figure}

Evidence for weak ferromagnetism has been reported in the EDO
class of \tp materials\cite{mur1,yosh1,yosh2,kon1,kon2} where
oxygen replaces the two nitrogen sites (N$_1$ and N$_2$ in Fig. 1)
in the DMEDT-TTF donor. Since there is no magnetic ion present,
magnetism must arise either from the bands or from disorder
(localized moments). Band magnetism has recently been treated by
Arita et al.\cite{ari}, based on consideration of the very flat,
narrow ($E_F \approx 8 meV$) bands that appear at the Fermi level
from the tight binding calculations. Although magnetization
studies have shown a small (0.001 $\mu_B$/formula unit) moment, no
hysteresis has been observed in magnetization data\cite{kon1}.
Recent far infrared measurements indicate that the methyl groups
show disorder, and it has been suggested that this may lead to
electronic localization and magnetism at low
temperatures\cite{ole}.

One final important detail is the relative constancy of the SdH
oscillation frequency in the vicinity of $B_{th}$, which indicates
that there is no catastrophic change in the quasi-two dimensional
(Q2D) electronic structure. Rather, the SdH waveforms (which
appear only below about 1.5 K due to the large effective
mass\cite{sto}) remain observable until the resistance becomes
forbiddingly large, either for increasing, or decreasing field
through \bt. These effects are particularly evident in Figs. 4b
and 5.

Based on the evidence above, we argue that the phenomena in the
\tp materials is linked to the disorder in the donor and anion
structure. The influence of high magnetic fields on this disorder
drives the system into a state of highly de-coupled two
dimensional Fermi liquid layers. This conclusion is supported by
noting that \bt~ is the lowest in the most disordered (racemic)
system. That the disorder, and the resulting incoherent transport
is limited to the inter-planer layers, is evident from the
presence of the SdH effect, and the absence (or attenuation) of
semiclassical orbits (AMRO) in tilted magnetic fields. Hence it is
unlikely that the mechanism that stabilizes the high field phase
is a result of Fermi surface nesting, especially in light of the
weak dependence of \bt~ on magnetic field direction. When the
field is aligned in-plane, \bt~ actually reaches a minimum, and
this may be due to further compression of the Q2D electron wave
function expected for a quasi-two-dimensional conductor in this
configuration.

A remaining question is how the magnetic field couples to the
inter-planar disorder, and we may speculate about two
possibilities. The first is simply through localized spins, where
in the low field phase, magnetic polarization may reduce carrier
scattering. This would give a negative magnetoresistance and allow
the SdH effect to eventually appear. In the high field phase,
magnetic order may set in, thereby making the inter-layer
transport incoherent due to first order, domain-like structures. A
second possibility may be diamagnetic in origin, arising from
non-stoichiometric sites in the inter-layer anion structure. The
high magnetic fields needed to cross \bt, and the linear shape of
the anions make (normally weak) diamagnetic processes, that may
alter the inter-layer structure, more probable. Either mechanism
should be roughly isotropic with respect to magnetic field
direction.

In summary, the similarities between organic molecular solids and
the metal oxides (cuprates, manganites, ruthenates, etc.)
continues to be the subject of general interest in materials
science. These relationships may be even closer, as evidenced from
the results presented in this Letter. Here, superimposed on the
``conventional" picture of a quasi-two dimensional Fermi surface,
we find evidence for phase transitions driven by disorder and
magnetism.

\begin{acknowledgments}
The FSU acknowleges support from NSF-DMR 99-71474. We thank B.
Ward, L. Gor'kov, E. Dagotto, and R. McKenzie for valuable
comments and suggestions.
\end{acknowledgments}



\begin{references}
\bibitem{cmr} For a review see E. Dagotto, T. Hotta, and A. Moreo,  Phys. Rep. {\bf 344}, 1 (2001).
\bibitem{tokura} Y. Tokura and Y. Tomioka, J. Mag. Mag. Mat.
\textbf{200}, 1(1999).
\bibitem{welp} U. Welp \emph{et al.}, Phys. Rev. Lett. \textbf{69},
840(1992).
\bibitem{shus} Y.V. Sushko \emph{et al.}, J. Superconductivity, \textbf{7}, 937(1994).
\bibitem{ross} R.H. McKenzie, Science, \textbf{278}, 820(1997).
\bibitem{elbio} E. Dagotto, Science, \textbf{293}, 2410(2001).
\bibitem{cao} G. Cao \emph{et al.}, Bull. Am. Phys.
Soc. \textbf{47}, 912(2002).

\bibitem{pap90}G. C. Papavassiliou et al., Organic Superconductivity
(Plenum Press, New York, 1990).
\bibitem{ter}A. Terzis \emph{et al.}, Synth.
Met. \textbf{42}, 1715 (1991).
\bibitem{zam}J. S. Zambounis \emph{et al.}, Solid State Commun. \textbf{95},
211 (1995).
\bibitem{pap95}G. C. Papavassiliou \emph{et al.},
Synthetic Metals \textbf{70}, 787 (1995).

\bibitem{pap96} G. C. Papavassiliou \emph{et al.}, Mol. Cryst. Liq. Cryst. \textbf{83}, 285 (1996).

\bibitem{ish}T.Ishiguro, K. Yamaji, and G. Saito, Organic Superconductors II
(Springer-Verlag, Berlin, Heidelberg, New York, 1998).
\bibitem{osada} T. Osada, H. Nose, M. Kuraguchi, Physica B
{\bf294-295}, 402(2001).
\bibitem{will} J. M. Williams \emph{et al.}, \emph{Organic Superconductors
(including Fullerenes)}, Prentice-Hall, Englewoods Cliffs, New
Jersey (1992).

\bibitem{pap01a}G. C. Papavassiliou \emph{et al.}, Synth. Met. \textbf{120}, 743 (2001).
\bibitem{pap01b}G.C. Papavassiliou, A. Terzis, and C. P. Raptopoulou, Naturforsch. {\bf 56b} , 963(2001).


\bibitem{sto} K. Storr \emph{et al.}, Phys. Rev. B \textbf{64}, 045107/1 (2001).

\bibitem{for}N. A. Fortune et al., Mat. Res. Soc. Symp. Proc. \textbf{328}, 307(1994).

\bibitem{bal}L. Balicas \emph{et al.}, Phys. Rev. Lett. \textbf{87}, 067002(2001).
\bibitem{wos}J. Wosnitza \emph{et al.}, Phys. Rev. Lett.\textbf{ 86}, 508(2001).

\bibitem{bro} J. S. Brooks \emph{et al.}, Mol. Cryst. Liq. Cryst. , in
press, and cond-mat/0012291(2002).

\bibitem{mur1} K. Murata \emph{et al.}, Synthetic Metals \textbf{86}, 2021 (1997).
\bibitem{yosh1}H. Yoshino \emph{et al.}, J. Phys. Soc. Japan \textbf{66}, 2410 (1997).
\bibitem{yosh2} H. Yoshino \emph{et al.}, J. Phys.
Soc. Japan \textbf{68}, 177 (1999).

\bibitem{kon1}K. Konoike \emph{et al.}, Synth. Met.
\textbf{120}(1-3),801(2001).
\bibitem{kon2}T. Konoike \emph{et al.}, J. Phys. Chem.
Solids in press (2002).

\bibitem{ari} R. Arita, K. Kuroki, and H. Aoki, Phys. Rev. B \textbf{61}, 3207 (2000).

\bibitem{ole}I. Olejniczak \emph{et al.}, Phys. Rev. B {\bf62} 15634(2000).







\end{references}
\end{document}